\documentstyle[version2,aps]{revtex}
\begin{document}
\draft
\twocolumn
\widetext
\begin{title}
Bogoliubov quasiparticle spectra of the effective $d$-wave model \\
for cuprate superconductivity
\end{title}
\author{Y. Ohta and M. Yamaguchi}
\begin{instit}
Department of Physics, Chiba University, Inage-ku, Chiba 263, 
Japan
\end{instit}
\author{R. Eder}
\begin{instit}
Department of Applied and Solid State Physics, 
University of Groningen, 9747AG Groningen, The Netherlands
\end{instit}
\receipt{\today}
\begin{abstract}
An exact-diagonalization technique on finite-size clusters is used 
to study the ground state and excitation spectra of the two-dimensional 
effective fermion model, a fictious model of hole quasiparticles derived 
from numerical studies of the two-dimensional $t$$-$$J$ model at low 
doping.  We show that there is actually a reasonable range of parameter 
values where the $d_{x^2-y^2}$-wave pairing of holes occurs and the 
low-lying excitation can be described by the picture of Bogoliubov 
quasiparticles in the BCS pairing theory.  The gap parameter of a size 
$\Delta_d$$\simeq$$0.13|V|$ (where $V$ is the attractive interaction 
between holes) is estimated at low doping levels.  The paired state 
gives way to the state of clustering of holes for some stronger 
attractions.  
\end{abstract}
\pacs{74.20.Mn, 71.27.+a, 71.10.+x}

\narrowtext
\topskip11.5cm

\section{INTRODUCTION}

High-temperature superconductors are characteristic of the very short 
coherence length and small carrier number, and may be in an intermediate 
regime between two limits, a BCS superconductor and a condensate of 
pre-formed bosons, and this observation may provide a possible way 
of explaining the so-called pseudo-gap (or spin-gap) anomaly commonly 
observed in the lightly doped regime of cuprate superconductors\cite{[1]}.  
Analyses of the two-dimensional (2D) $t$$-$$J$ model at low doping 
have suggested a possible picture\cite{[2],[3],[4]};  under strong 
antiferromagnetic spin correlations, low-energy properties may be described 
by the coherent drift-motion of the spin-bag--like quasiparticles\cite{[5]} 
where the rapid incoherent oscillation of the `bare hole' is eliminated 
as excitations inside the bag\cite{[3]}.  This picture immediately suggests 
an effective fermion model, i.e., a fictious system of interacting fermions 
representing the hole quasiparticles\cite{[6],[7],[8]}.  
Since the attractive interaction acting between hole quasiparticles 
(or binding energy) is fairly strong $\sim$$0.8J$ in comparison with the 
quasiparticle bandwidth $\sim$$2J$ \cite{[2],[9]}, one may hope that the 
effective model simulating this situation would provide a possible 
explanation of some of the anomalous quasiparticle properties of 
cuprate superconductors.  

The purpose of this paper is then to present an analysis of the ground 
states and some excitation spectra of finite-size clusters of the effective 
fermion model at $T$$=$$0$ K.  We show that there is actually a reasonable 
range of parameter values where the $d$-wave pairing of holes occurs and 
the low-lying excitation can be described by the picture of Bogoliubov 
quasiparticles in the BCS pairing theory\cite{[10]}.  Also we show that 
the $d$-wave gap parameter is nearly in proportion to but much smaller 
than the strength of the attractive interaction between holes.  However, 
clustering of holes occurs for some stronger attractions, unlike, e.g., 
in the negative-$U$ Hubbard model where the coherence length continuously 
decreases with the attraction strength and a condensate of tightly-bound 
real-space pairs becomes a good description\cite{[11]}.   

\topskip0cm

The effective fermion model considered here may be defined by the 
Hamiltonian 
\begin{equation}
H=\sum_{{\bf k}\sigma} \varepsilon_{\bf k} 
c_{{\bf k}\sigma}^\dagger c_{{\bf k}\sigma}
+ V \sum_{<ij>} n_i n_j
\label{eq1}
\end{equation}
with a negative value of the density interaction $V$ acting on holes in 
the nearest-neighbor sites $<$$ij$$>$\cite{[6],[7],[8]}.  
$n_i$ is the number operator at site $i$, and $c^\dagger_{{\bf k}\sigma}$ 
creates a hole (`spinless' fermion) with momentum ${\bf k}$ in the $\sigma$ 
($=$$\uparrow$, $\downarrow$) sublattice; we have the two-sublattice model 
where the holes with up-spin (down-spin) are always in the 
$\uparrow$-sublattice ($\downarrow$-sublattice) and the numbers of up-spin 
and down-spin holes are the same.  The single-hole dispersion in the 2D 
antiferromagnet is taken as the noninteracting band structure 
$$
\varepsilon_{\bf k}=4t_{11}\cos k_x \cos k_y 
+ 2t_{20}(\cos 2k_x + \cos 2k_y)
$$
where $t_{11}$ and $t_{20}$ are the second- and third-neighbor hopping 
integrals with positive sign.  We employ the Lanczos exact-diagonalization 
technique on finite-size clusters of up to 64 sites with periodic boundary 
condition; we can take clusters much larger in size than those feasible 
for the $t$$-$$J$ model.  Hereafter we refere to the momentum defined with 
respect to the `nonmagnetic' Brillouin zone (see Fig.~1).

\section{GROUND STATES}

The ground states with zero total momentum are calculated as a function 
of the parameter values $t_{20}/t_{11}$ and $V/t_{11}$ for all the 
clusters with 32, 36, 50, and 64 sites.  We find that, irrespective of 
the size of the clusters, the ground states have the point-group symmetries 
in the parameter space, as those shown in Fig.~2; there are some quantitative 
differences in the phase boundary among clusters but the basic features 
are the same.  

First, we note that two holes in the empty lattice form a $d_{x^2-y^2}$-wave 
bound pair for $t_{20}/t_{11}$$\lesssim$$0.6$$-$$0.8$; this is simply because 
the two holes can gain kinetic energy only by changing the spatial sign 
of the pair wave-function since a hole goes round the other with the 
positive hopping parameter $t_{11}$\cite{[8]}.  
For $t_{20}/t_{11}$$\gtrsim$$0.6$$-$$0.8$, this mechanism of $d$-wave 
pairing works less favorably, and also in the ${\bf k}$-space, holes 
tend to accumulate around ${\bf k}$$=$$(\pm\pi/2,\pm\pi/2)$ where there 
is a node of the $d$-wave pair wave-function, so that the $p$-wave 
pairing occurs.  
There are two regions in the four-hole case; at $t_{20}/t_{11}$$\lesssim$$0.6$ 
and $|V|/t_{11}$$\lesssim$$4$ we have the region where there are two 
$d$-wave pairs (as will be shown in Sec.~IV).  In the six-hole case we 
again find a parameter region where three $d$-wave pairs present.  
Clustering of holes occurs however for larger attraction strength, 
suggesting the model to be phase separated.  
This is evident in the calculated density correlation function 
$\langle$$n_in_j$$\rangle$ where $\langle$$...$$\rangle$ denotes the 
ground-state expectation value; we find that with increasing $|V|/t_{11}$ 
the probability of finding a hole in the nearest and next-nearest 
neighbors of a site with a hole increases suddenly across the 
level-crossing line in Fig.~2 and saturates rapidly to a constant value, 
and it also increases very rapidly around $V/t_{11}$$\simeq$$-$$2$ 
even for $t_{20}/t_{11}$$\gtrsim$$0.5$.  

We should note here that in the effective model the hole can jump over 
the other because of the hopping term to the second and third neighbors.  
This process however is forbidden in the $t$$-$$J$ model.  The stronger 
tendency to hole clustering in the effective model may therefore be 
suppressed by somehow implementing a `conditional hopping' of the $t$$-$$J$ 
model so as to gain in kinetic energy of holes.  Such a correction however 
will be discussed elsewhere.  Hereafter, we present our analysis of the 
model Eq.~(1) in the parameter region where the clustering of holes does 
not occur (i.e., smaller $|V|/t_{11}$ values) and two holes in the empty 
lattice form a $d$-wave bound state (i.e., $t_{20}/t_{11}$$\lesssim$$0.5$).  
We thereby examine what are the low-energy excitations in this parameter 
region.

\section{SINGLE-PARTICLE SPECTRA}

We first calculate the single-particle spectral function defined as 
$A({\bf k},\omega)=A^-({\bf k},-\omega)+A^+({\bf k},\omega)$
with
\begin{eqnarray}
&A&^-({\bf k},\omega)=\sum_{\nu\sigma}|\langle\psi^{N-1}_\nu |
c_{{\bf k}\sigma}|\psi^N_0\rangle |^2
\delta(\omega-E^{N-1}_\nu+E^N_0)
\nonumber \\
&A&^+({\bf k},\omega)=\sum_{\nu\sigma}|\langle\psi^{N+1}_\nu |
c^\dagger_{{\bf k}\sigma}|\psi^N_0\rangle |^2
\delta(\omega-E^{N+1}_\nu+E^N_0)
\nonumber \\
\end{eqnarray}
which simulates the angle-resolved photoemission (PES) and 
inverse photoemission (IPES) spectroscopy.  
$E^N_\nu$ and $\psi^N_\nu$ are the $\nu$-th excited eigenenergy 
and eigenstate ($\nu$$=$$0$ denotes the ground state) of the 
$N$-hole system of the cluster, respectively.  We add a small 
imaginary number $i\eta$ to $\omega$ and give a Lorentzian broadening 
to the spectra which otherwise consist of a set of $\delta$ 
functions; the value $\eta/t_{11}$$=$$0.05$ is used.  
This spectra can in principle be compared to recent experimental 
data\cite{[1]}.  

The calculated results for $A({\bf k},\omega)$ are given in Fig.~3.  
When $|V|/t_{11}$ value is small the spectra simply reflect 
the noninteracting band structure, but for larger values of $|V|/t_{11}$ 
we find that the spectra are sharp near the Fermi energy but become 
broad away from the Fermi energy and momentum, reminiscent of the 
`dressing' of the particle of correlated Fermi-liquid systems; 
the Bloch electron added into the correlated Fermi sea may be 
scattered out of its single electron level, leaving the system 
in the manifold of excited states.  

The calculated spectra also show how the attractive interaction affects 
the single-particle excitation spectrum.  The gap-like structures can 
be seen in the spectra at (and around) ${\bf k}_{\rm F}$, and the size 
of the gap increases with increasing attraction strength.  The spectral 
feature, i.e., spectral-weight transfer appearing in both PES and IPES 
sides near ${\bf k}_{\rm F}$ and vanishing far away from ${\bf k}_{\rm F}$, 
is noticed, which is consistent with the expectation of the BCS theory.  
This feature is absent along the diagonal ($k_x$$=$$k_y$) of the Brillouin 
zone and is consistent with the $d_{x^2-y^2}$-wave pairing.  
The spectral-weight transfer also indicates a tendency toward smearing 
of the jump at ${\bf k}_{\rm F}$ in the momentum distribution 
function.

\section{BOGOLIUBOV QUASIPARTICLES}

Now let us examine the validity of the Bogoliubov-quasiparticle 
picture for the low-lying states of the effective model.  
We use the technique proposed in Ref.\cite{[12]}, i.e., an exact 
calculation of Bogoliubov quasiparticle spectrum on small clusters, 
and see if this picture works in this model.  
We define the one-particle anomalous Green's function as 
\begin{equation}
G({\bf k},z)=
\langle\psi^{N+2}_0|c^\dagger_{{\bf k}\uparrow} 
{1 \over {z-H+E_0}}
c^\dagger_{-{\bf k}\downarrow}|\psi^{N}_0\rangle
\label{eq3}
\end{equation}
with the spectral function 
$$
F({\bf k},\omega)=-{1\over \pi}{\rm Im}G({\bf k},\omega+i\eta)
$$
and its frequency integral 
$$
F_{\bf k}=\langle 
\psi^{N+2}_0|c^\dagger_{{\bf k}\uparrow}
c^\dagger_{-{\bf k}\downarrow}|\psi^N_0\rangle. 
$$
where $E_0$ is chosen as the average of $E_0^N$ and $E_0^{N+2}$.  
The hypothesis (see Ref.\cite{[12]} for details) that low-lying states 
of the clusters can be described by the microcanonical version of the 
BCS pairing theory then predicts that 
$F({\bf k},\omega)$$=$$F_{\bf k}\delta (\omega-E_{\bf k})$ 
with $F_{\bf k}$$=$$\Delta_{\bf k}/2E_{\bf k}$, 
where $E_{\bf k}$ and $\Delta_{\bf k}$ are the quasiparticle 
energy and gap function, respectively.  Similarly, we define the 
two-particle anomalous Green's function as 
\begin{eqnarray}
G({\bf k},{\bf k'},z)&=&
\langle\psi^N_0|
c^\dagger_{-{\bf k'}\downarrow}c_{-{\bf k}\downarrow}
{1\over{z-H+E^N_0}}
c^\dagger_{{\bf k'}\uparrow}c_{{\bf k}\uparrow}
|\psi^N_0\rangle
\nonumber \\
&\;&\;\;\;\;\;
-{n_{-{\bf k}\downarrow}n_{{\bf k}\uparrow} \over {z}}
\delta_{\bf kk'} 
\label{eq4}
\end{eqnarray}
with $n_{{\bf k}\sigma}$$=$$\langle\psi^N_0| 
c^\dagger_{{\bf k}\sigma} c_{{\bf k}\sigma}| 
\psi^N_0\rangle$.  
Only the $N$-particle subspace is involved here unlike in 
Eq.~(\ref{eq3}).  We define the spectral function 
$F({\bf k},{\bf k'},\omega)$ and its frequency integral 
$F_{\bf kk'}$ as above.  The hypothesis then predicts 
$F({\bf k},{\bf k'},\omega)$$=$$F_{\bf k} 
F_{\bf k'}^* \delta(\omega$$-$$E_{\bf k}$$-$$E_{\bf k'})$. 
By examining these two anomalous Green's functions, we can see if 
the low-energy excitations of the model are described by the BCS 
pairing theory.  

The spectral function $F({\bf k},\omega)$ is calculated for a 
64-site cluster of the effective model and is shown in Fig.~4 
for a number of the attraction strength.  
Noticing that the calculation is made by the subtraction of two 
single-particle excitation spectra\cite{[12]}, we should first of all 
stress that nearly all incoherent spectral features in 
$A({\bf k},\omega)$ are subtracted out, leaving only the low-energy 
features (stemming from weakly-interacting Bogoliubov 
quasiparticles as shown below). 

We then find the following, all of which are consistent with 
expectations of the BCS pairing theory:  
A pronounced low-energy peak appears at ${\bf k}_{\rm F}$ and smaller 
peaks appear at higher energies for other momenta; the weights of the 
peaks are consistent with the BCS form of the condensation ampitude 
$F_{\bf k}$ with a maximum at ${\bf k}_{\rm F}$ (see Fig.~5).  The 
momentum dependence of $F({\bf k},\omega)$, i.e., the change in sign 
under rotation by $\pi/2$ and vanishing weight along the $k_x$$=$$k_y$ 
line, is a clear indication of $d_{x^2-y^2}$-wave pairing.  The size of 
the energy gap, which may be estimated directly from the positions of 
the peaks, increases with increasing $|V|/t_{11}$ value.  With decreasing 
gap size, the peaks at momenta other than ${\bf k}_{\rm F}$ lose their 
weight as expected from the BCS theory.  We note that, when the 
attraction strength $V/t_{11}$ exceeds some critical value, these 
features are collapsed and the spectra completely loose the significance 
of Bogoliubov quasiparticles, indicating that the clustering of holes 
takes place and the system seems to phase separate.  
The coherence length $\xi$ estimated from the momentum dependence 
of $F_{\bf k}$ decreases with the attraction strength but still 
is more than twice as large as the nearest-neighbor lattice spacing 
even near the critical $V/t_{11}$ value.  

We further compare the calculated spectra $F({\bf k},\omega)$ with 
the BCS predictions where we assume the quasiparticle energy 
$E_{\bf k}$$=$$\sqrt{(\varepsilon_{\bf k}-\mu)^2+\Delta_{\bf k}^2}$ 
with the gap function $\Delta_{\bf k}$$=$$\Delta_d (\cos k_x-\cos k_y)$ 
and choose the chemical potential $\mu$ to guarantee vanishing 
$\varepsilon_{\bf k}$$-$$\mu$ at ${\bf k}_{\rm F}$.  The value of 
$\Delta_d$ is then evaluated by fitting the positions of the low-energy 
peaks in $F({\bf k},\omega)$.  We find that the fitted quasiparticle 
spectra are in a fair agreement with the exact spectra, which 
demonstrates the validity of the BCS pairing theory for low-lying 
excitations in the effective model.  The estimated values of the gap 
parameter $\Delta_d$ are shown in Fig.~\ref{fig6}.  We find that 
$\Delta_d$ is insensitive to the doping levels examined and has the 
value $\Delta_d$$\simeq$$0.13|V|$ until reaching the region of phase 
separation, which is much smaller than the bandwidth for 
$\varepsilon_{\bf k}$.  This value also corresponds to 
$\Delta_d$$\sim$$130$\,K if we assume $|V|$$\sim$$0.8J$ and 
$J$$\sim$$1300$\,K\cite{[9]}.  

A consequence of $d_{x^2-y^2}$-wave pairing may be seen in the 
point-group symmetry of the ground states, i.e., an alternation of 
the symmetry between $B_1$ and $A_1$ for the two-, four-, and six-hole 
states in a parameter region (see Fig.~2).  
This alternation, present also in the $t$$-$$J$ clusters\cite{[12]}, 
suggests the picture that holes are added in pairs with 
$d_{x^2-y^2}$-wave symmetry.  $G({\bf k},z)$ in Eq.~(\ref{eq3}) 
(and thus $F({\bf k},\omega)$) reflects the pairing symmetry via 
the point-group symmetries of $\psi^N_0$ and $\psi^{N+2}_0$.  
Note that $F({\bf k},{\bf k'},\omega)$ is defined entirely within 
the $N$-particle subspace and thus is not affected by this alternation; 
nevertheless it shows the same indication of $d_{x^2-y^2}$-wave pairing 
as $F({\bf k},\omega)$.  
The calculated results for $F({\bf k},{\bf k'},\omega)$ 
are shown in Fig.~\ref{fig7}.  We again find that the size of 
the excitation gap increases with increasing $|V|/t_{11}$, and the 
${\bf k}$-dependence of the spectra clearly indicates 
$d_{x^2-y^2}$-wave pairing.  
There are sharp peaks at low energies and broadened features at higher 
energies.  As is the case for $F({\bf k},\omega)$, these high-energy 
features lose their weight rapidly with decreasing $|V|/t_{11}$ value, 
whereas the peaks at ${\bf k}_{\rm F}$ become sharp but remain finite 
with decreasing $|V|/t_{11}$ value.  These results are consistent with 
the notion of weakly-interacting Bogoliubov quasiparticles for 
low-lying excitations in the BCS superconductors.  When the attraction 
strength $V/t_{11}$ exceeds some critical value, the spectra 
$F({\bf k},{\bf k'},\omega)$ again completely loose the significance 
of Bogoliubov quasiparticles, indicating that the clustering of holes 
takes place.  

\section{SUMMARY}

By using an exact-diagonalization technique on finite-size 
clusters, we have studied the ground states and excitation spectra 
of the 2D effective fermion model, i.e., a system of spin-bag--like 
quasiparticles corresponding to doping holes, which is derived 
from numerical studies of the 2D $t$$-$$J$ model at low doping.  
We have examined the point-group symmetries and density correlations of 
the ground states of various size clusters with two, four, and six holes 
over a wide range of parameter values, and have found that the results 
are insensitive to the size of the clusters and can be summarized as 
the schematic phase diagram reflecting the pairing symmetry of holes.  
We then have calculated the single-particle spectra of the model in 
a parameter region and have detected some indications of the 
$d_{x^2-y^2}$-wave hole pairing in the obtained spectral features.  
We have further calculated the one-particle and two-particle anomalous 
Green's functions and found that there is actually a reasonable range 
of parameter values where the $d_{x^2-y^2}$-wave pairing of holes occurs 
and the low-lying excitation can be described by the picture of 
Bogoliubov quasiparticles in the BCS pairing theory.  The gap parameter 
is estimated to be $\Delta_d$$\simeq$$0.13|V|$ at low doping levels.  
It seems however that with increasing the attraction strength the 
state of clustering of holes overwhelms this $d_{x^2-y^2}$-wave 
pairing state before the picture of a condensate of tightly-bound 
pairs with the $d_{x^2-y^2}$-wave internal structure becomes 
appropriate.  

\acknowledgments
This work was supported in part by Grant-in-Aid for Scientific Research 
from the Ministry of Education, Science, and Culture of Japan.  
Financial support of Y.O. by Saneyoshi Foundation and R.E. by European 
Community is gratefully acknowledged.  Computations were carried out in 
the Computer Centers of the Institute for Solid State Physics, University 
of Tokyo, and the Institute for Molecular Science, Okazaki National 
Research Organization.

\figure{Available ${\bf k}$-points in the Brillouin zone for 
(a) 32-, (b) 36-, (c) 50-, and (d) 64-site clusters with periodic 
boundary condition.  
The dotted lines indicate the Brillouin zone defined for the 
two-sublattice model.  
\label{fig1}}

\figure{Schematic representation of the point-group symmetry 
of the ground state at ${\bf k}$$=$$(0,0)$ obtained for the 32-, 36-, 50-, 
and 64-site clusters with (a) two holes, (b) four holes, 
and (c) six holes; B$_1$ ($d_{x^2-y^2}$), E ($p_x$ and $p_y$), 
and A$_1$ ($s$) symmetries appear in the parameter space.  
\label{fig2}}

\figure{Single-particle spectra $A({\bf k},\omega)$ calculated for 
various attraction strengths ($V/t$$=$$-$$1$, $-$$2$, and $-$$3$ for 
left, center, and right columns, respectively) and hole numbers (two, 
four, and six holes).  $\omega$ is the energy of the hole.  
The spectra for the 64-site cluster are arranged from top to bottom 
in each panel as ${\bf k}$$=$$(\pi/2,\pi/2)$, $(\pi/4,\pi/4)$, $(0,0)$, 
$(\pi/4,0)$, $(\pi/2)$, $(3\pi/4,0)$, $(\pi,0)$, $(3\pi/4,\pi/4)$, 
and $(\pi/2,\pi/4)$.  Those for the 32-site cluster are as 
${\bf k}$$=$$(\pi/2,\pi/2)$, $(\pi/4,\pi/4)$, $(0,0)$, 
$(\pi/2,0)$, $(\pi,0)$, and $(3\pi/4,\pi/4)$.  
\label{fig3}}

\figure{Bogoliubov quasiparticle spectra $F({\bf k},\omega)$ for a 
number of attraction strengths.  The upper four panels show the 
results from the anomalous Green's function between two- and four-hole 
ground states for the 64-site cluster with $t_{20}/t_{11}$$=$$0.3$, and 
the lower four panels show the results between four- and six-hole 
ground states for the 32-site cluster with $t_{20}/t_{11}$$=$$0.3$.  
The arrangements of the spectra in each panel follow those in Fig.~3.  
\label{fig4}}

\figure{Attraction strength dependence of the condensation amplitude 
$F_{\bf k}$ calculated for the 64-site cluster with the filling 
between two and four holes (left column) and for the 32-site cluster 
between four and six holes (right column).  
\label{fig5}}

\figure{Gap parameter $\Delta_d/t_{11}$ as a function of the attraction 
strength $|V|/t_{11}$ estimated from the calculated $F({\bf k},\omega)$ 
for two-, four-, and six-hole systems.  
\label{fig6}}

\figure{Bogoliubov quasiparticle spectra $F({\bf k},{\bf k}',\omega)$ 
calculated for the 32-site cluster with six holes.  The spectra for 
various ${\bf k}$ points with ${\bf k}'$$=$$(3\pi/4,-\pi/4)$ are shown.  
$t_{20}/t_{11}$$=$$0.3$ and $V/t_{11}$$=$$-$$2.0$ are used.  
\label{fig7}}

\end{document}